\documentclass[aps,amsmath,twocolumn,amssymb,floatfixng,showpacs,superscriptaddress,footinbib]{revtex4-1}
\pdfoutput=1
\usepackage[dvips]{graphics}
\usepackage{tikz}
\usepackage{bm}
\usepackage{float}
\usepackage{epsfig}
\usepackage{enumerate}
\usepackage{subfigure}
\usepackage{amsmath}
\usepackage{color}
\usepackage{braket}
\usepackage{graphicx}
\usepackage{float}
\usepackage[colorlinks=true,linktoc=page,linkcolor=red,citecolor=blue,urlcolor=magenta]{hyperref}

\newcommand\bea{\begin{eqnarray}}
\newcommand\eea{\end{eqnarray}}
\newcommand\beq{\begin{equation}}  
\newcommand\eeq{\end{equation}}

\begin{document} 
\title{Intriguing topological superconductor phases in long-range extended Kitaev models: an interplay between sub-lattices and  power-law interaction} 
 \author{Dharana Joshi}
\affiliation{Department of Physics, BITS Pilani-Hyderabad Campus, Telangana 500078, India}
\author{Amaan Khan}
\affiliation{Department of Physics, BITS Pilani-Hyderabad Campus, Telangana 500078, India}
\affiliation{Microsoft Campus,  Gachibowli, Hyderabad, Telangana 500032, India}
\author{Tanay Nag}
\email{tanay.nag@hyderabad.bits-pilani.ac.in}
\affiliation{Department of Physics, BITS Pilani-Hyderabad Campus, Telangana 500078, India}


\begin{abstract}

The power-law profile of hopping and superconductivity introduces rich critical phenomena that are not possible to explore in the nearest-neighbor models. Having this in mind, we construct the 
Su–Schrieffer–Heeger (SSH)-Kitaev model with mono-, bi-, and tri-partite sub-lattices where superconductivity [hopping] is of power-law Kitaev [SSH] type. The long-range (LR) superconductivity can alone gap out the zero-momentum critical line while LR hopping  alters the finite-momentum critical line. There exists open-ended critical line at vanishing superconducting gap on the phase diagram for all versions of the SSH-Kitaev model. Interestingly, LR superconductivity (hopping) promotes massive Dirac (Majorana zero) modes, while in a finite-size system, their emergence in open boundary conditions is caused by the bulk gap of the anti-periodic Hamiltonian in momentum space. This indicates an unconventional bulk-boundary correspondence for power-law models as compared to what is seen in the short-range models. Importantly, the thermodynamic phase boundaries are correctly reproduced by the above unconventional correspondence. Intriguingly, an even (odd) number of sub-lattices yields an integer (half-integer) winding number while the  
 massive Dirac modes continue to survive irrespective of the sub-lattice profile as long as there exists LR pairing only.

\end{abstract}

\maketitle

\textcolor{blue}{\textit{Introduction}}--In the class of quantum materials, 
topological superconductors have attracted considerable interest due to their unconventional quantum phases, hosting Majorana zero modes (MZMs), and subsequently providing a promising platform for error-resistant
quantum computation and various
potential applications in quantum information science \cite{Deng_2012,Mourik_2012,Albrecht_2016,Nadj_Perge_2014,Sato_2017,Hansson_2012,Ivanov,KITAEV20032,Stern2010,CNayak}. 
The Kitaev chain \cite{Kitaev_2001} represents a one-dimensional paradigmatic model for studying topological superconductivity where Cooper pairs form from spinless fermions with $p$-wave pairing rather than conventional $s$-wave pairing of electrons with opposite spins.  Interestingly,  MZMs, residing at the  system's boundaries, lie within the superconducting gap and, unlike conventional superconductors, are topologically protected against local perturbations \cite{Zhang_2025,Kitaev_2001} and are characterized by quantized winding number \cite{Song2019,He_2021,joshi2025A,Lin2021}. 
However, due to the lack of experimental realization of $p$-wave superconducting gap, Rashba 
nanowire model comes out as the most promising and experimentally viable platforms for realizing topological superconductivity \cite{Lutchyn_Sau, Mourik,SAU_PRL, PRL_SAU,Beenakker}. Nevertheless, the $p$-wave Kitaev chain remains serving as a generic as well as minimal model to explore various quantum effects \cite{PhysRevB.106.054308,PhysRevB.107.184311,Mondal2024,Anirban2017,PhysRevE.90.042107,PhysRevLett.110.146404,PhysRevB.88.165111,PhysRevE.89.042125,PhysRevE.96.022136,PhysRevB.88.155133,PhysRevB.110.064302,PhysRevB.106.L220506,he2024real,PhysRevB.105.085106}.

Recent developments in atomic, molecular, and optical platforms have enabled the power-law interactions to be realized in experiments \cite{PhysRevLett.104.240403,PhysRevLett.113.030602,richerme2014non,britton2012engineered,PhysRevLett.121.093602}. This leads to a growing interest in exploring long-range (LR) models with a power law $1/r^{\alpha}$-profile where $r$ denotes the distance \cite{PhysRevB.98.014204,PhysRevB.99.224203,PhysRevB.99.104203}.  Such models in the presence of on-site disorder can exhibit algebraically localized states, in contrast to the conventional exponentially localized states found in short-range (SR) disordered systems \cite{PhysRevB.95.094205,PhysRevResearch.2.012074,PhysRevE.101.052108}. Similar to the fermionic and spin models,  the LR superconducting pairing in the Kitaev model exhibit interesting phases and critical behavior as compared to the regular Kitaev model with nearest-neighbour $p$-wave pairing \cite{PhysRevB.110.064302,Anirban2017,Vodola2014,Vodola_2016,Lepori_2017,Bhattacharya_2019,Fraxanet2021}. To be precise, one of the existing phase boundaries dissolves under LR nature of the superconductivity and MZMs transform into massive Dirac modes (MDMs) \cite{Vodola2014,Vodola_2016,Lepori_2017,Bhattacharya_2019,Fraxanet2021,Solfanelli_2023,Cinnirella2025,Alecce2017,Gandhi2025,Gao2015,Fraxanet2021,Viyuela2016,haink2025,Baghran_2024,Mondal_2022}. Notably, half-integer quantization of winding number is noticed for MDMs.
Furthermore, LR hopping can also be introduced, where the hopping amplitude between two sites separated by a distance $r$ decays as a power law $1/r^\epsilon$, leads to additional nontrivial topological phases and further modifications of the phase boundaries \cite{Alecce2017,Xue-Si2020}.

On the other hand, the Su–Schrieffer–Heeger (SSH) model provides a  well-established framework for understanding topological insulator properties in one-dimensional systems \cite{Anastasiadis22, Verma24, du2024one, dharana2025, rajbongshi2025topological, asboth2016short, Maffei_2018, Perez2019, Du_2024, Feng2022, Halder_2022, Lieu_2018, Nehra_2022, Yin18,hj3p-d7vl,Perez2019}. Note that $p$-wave Kitaev  and SSH chains both serve as paradigmatic models 
to establish the 
bulk–boundary correspondence (BBC) in one-dimension connecting the topological properties of a bulk system and the emergence of characteristic boundary states. Given the above background on $p$-wave Kitaev models and SSH model, it would be interesting to study the extended Kitaev models where the electronic part hosts SSH-type power-law hopping and superconductor part contains power-law $p$-wave pairing. This model allows us to explore the following  questions: How are the critical phenomena affected by power-law hopping and superconductivity?
How does the LR behavior affect the  BBC? What is the effect of the sublattice on the emergent topology? 

In this work, we study the SSH$n$-Kitaev chain, with $n$ representing the number of sublattices, and provide a systematic analysis of its topological superconductor phases for different power-law decay exponents. By tuning the exponent, we bridge the SR and LR regimes of the model, thereby revealing the evolution of topological phases and their boundaries across these limits as well as variations in sublattice profiles. Remarkably, in the LR pairing limit, the MDMs are characterized by a half-integer bulk winding number, evaluated for a finite system under open boundary condition (OBC), while changes in this topological invariant are associated with bulk gap closings in the corresponding system under anti-periodic boundary condition (ABC). This OBC-ABC  correspondence reduces to the regular BBC in the SR limit of pairing and hopping. Surprisingly, the half-unity winding number is noticed only for an odd value of $n$,   while for even $n$, the winding number approaches unity. Importantly, thermodynamic phase boundaries are correctly reproduced following the OBC-ABC correspondence that further confirms the robustness of our findings.



\begin{figure*}
    \centering
    \includegraphics[width=0.32\linewidth]{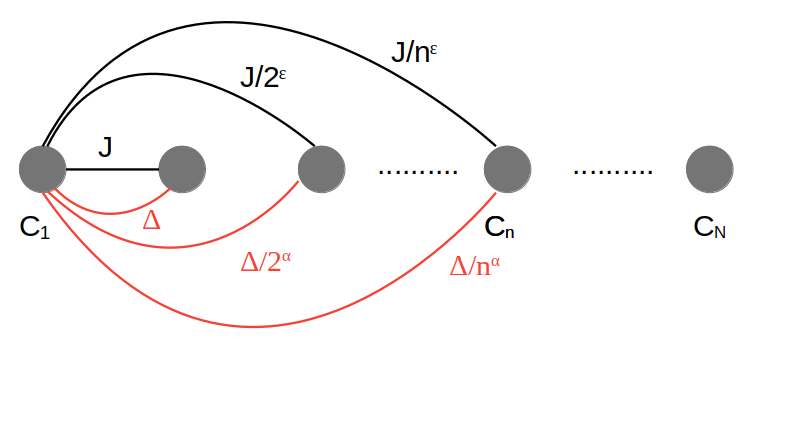}
    \includegraphics[width=0.32\linewidth]{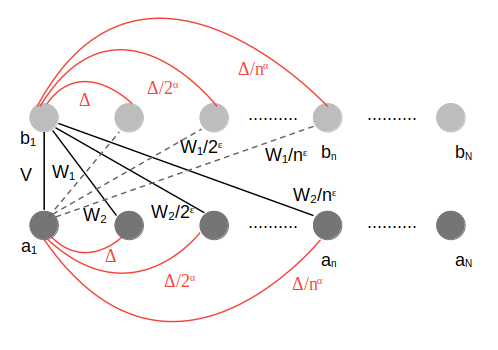}
    \includegraphics[width=0.32\linewidth]{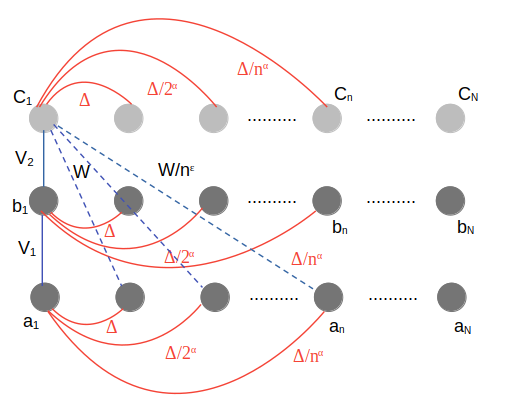}
    \caption{We schematically represent SSH1-Kitaev, SSH2-Kitaev and SSH3-Kitaev models, having one, two and three sub-lattices, respectively, in (a,b,c). 
    In (a), $J$ ($\Delta$) represents hopping (superconducting pairing) term, designated by black (red) color,   and $\epsilon$ $(\alpha)$ denotes hopping (superconducting) power-law exponent. In (b),  $V$ is intra-cell hopping, $W_1$ and $W_2$ are inter-cell hopping. In (c),  $V_1$ and $V_2$ are intra-cell hopping, and $W$ is inter-cell hopping.}
    \label{fig:Model1}
\end{figure*}

\textcolor{blue}{\textit{Model and invariant}}--We consider power-law  SSH1-, SSH2- and SSH3-Kitaev models with mono-, bi- and tri-partite structure of the sub-lattice, respectively. Note that the SSH1-Kitaev corresponds to the LR Kitaev model, shown in Fig. \ref{fig:Model1} (a),  as given below   
\begin{eqnarray}
    H_{1} &=& -\sum_{i=1}^{N'} \sum_{r=1}^{N_e}\left[ \frac{J}{r^\epsilon} d_i^\dagger d_{i+r}+ \frac{\Delta}{r^\alpha} d_i ^\dagger d_{i+r}^\dagger+h.c.\right] \nonumber \\
    &-& \mu  \sum_{i=1}^N \left(1-d_i^\dagger d_i \right ) 
\label{eq:eq1}
\end{eqnarray}
Here, $d_i^\dagger$ and $d_i$ are spinless fermionic creation and annihilation operators at site $i$. For OBC, $(N',N_e)=(N-1,N-1)$ and for PBC/ABC, $(N',N_e)=(N/2,N/2)$.   The first term
describes the kinetic part namely, hopping between sites, separated by distance $r$, with amplitude $J r^{-\epsilon}$. The second term represents the $p$-wave
superconducting pairing between sites, separated by distance $r$,  with strength $\Delta r^{-\alpha}$.  The last term represents the effect of on-site chemical potential $\mu$. As $\epsilon,\alpha \to \infty$, the above model effectively mimics a regular $p$-wave Kitaev chain, supporting topological superconductor phase for $-2<\mu/J<2$ and $\Delta \ne 0$. 

Proceeding with the SSH2-Kitaev model, the unit cell is now composed of two sub-lattice degrees of freedom such that $d \to (a,b)$ and $d^\dagger \to (a^\dagger,b^\dagger)$, see   Fig. \ref{fig:Model1} (b). The intra-cell hopping $V$ is spatially independent. The inter-cell hopping between $a_i$ ($b_i$) and $b_{i+r}$ ($a_{i+r}$) is given by  $W_1 r^{-\epsilon}$ ($W_2 r^{-\epsilon}$). The inter-cell  $p$-wave superconductivity is considered to be intra-orbital i.e., between $a_i$ and $a_{i+r}$ ($b_i$ and $b_{i+r}$) with same amplitude $\Delta r^{-\alpha}$. The real space Hamiltonian of the system  is given by
\begin{align}
H_{2}  &=\sum_{i=1}^{N} Va_i^\dagger b_i  + 
\sum_{i=1}^{N'} \sum_{r=1}^{N_e} \left(\frac{W_1}{r^\epsilon}a_i^\dagger b_{i+r}  + 
 \frac{W_2}{r^\epsilon}b_i^\dagger a_{i+r}  \right)+h.c.\nonumber \\  
 &+ \sum_{i=1}^{N'} \sum_{r=1}^{N_e} \frac{\Delta}{r^\alpha}\left(a_i^\dagger a_{i+r}^\dagger  +  b_i^\dagger b_{i+r}^\dagger + h.c.\right)\nonumber \\   &+ \mu  \sum_{i=1}^N \left(1-a_i^\dagger a_i -b_i^\dagger b_i\right )
\label{eq:realssh}
\end{align}

In the similar fashion, the SSH3-Kitaev model, defined for three sub-lattice degrees of freedom $(a,b,c)$, such that $d \to (a,b,c)$ and $d^\dagger \to (a^\dagger,b^\dagger,c^\dagger)$ is given by 
\begin{align}
H_{3} & =\sum_{i=1}^{N} V_1a_i^\dagger b_i + V_2b_i^\dagger c_i + 
\sum_{i=1}^{N'} \sum_{r=1}^{N_e} \frac{W}{r^\epsilon}a_i^\dagger c_{i+r} +h.c. \nonumber \\  
 &+ \sum_{i=1}^{N'} \sum_{r=1}^{N_e} \frac{\Delta}{r^\alpha}\left(a_i^\dagger a_{i+r}^\dagger + b_i^\dagger b_{i+r}^\dagger + c_i^\dagger c_{i+r}^\dagger + h.c.\right) \nonumber \\   &+ \mu  \sum_{i=1}^N \left(1-a_i^\dagger a_i -b_i^\dagger b_i - c_i^\dagger c_i\right )
  \label{eq:realssh3}
\end{align}
The intra-cell hopping $V_{1,2}$ is spatially independent. The inter-cell hopping between $a_i$ and $c_{i+r}$  is given by  $W r^{-\epsilon}$, see  Fig. \ref{fig:Model1} (c). Similar to SSH2-Kitaev model, the inter-cell  $p$-wave superconductivity is considered to be intra-orbital with identical amplitude $\Delta r^{-\alpha}$ for the orbitals.


Note that the ground state energy  $E$ for LR model with power-law terms can scale super-extensively in certain parameter regions. In order to obtain thermodynamically stable results, one can divide the model Hamiltonian by $N^{f(\alpha, \epsilon)}$ where $f(\alpha, \epsilon)>1$ [$=1$] for $\alpha<1,\epsilon<2$ [$\alpha<1$]. However, findings remain qualitatively the same irrespective of the factor $f(\alpha, \epsilon)$,  see SM \cite{supp} for more details.





We consider the winding number to be the appropriate topological invariant as the models have chiral symmetry $C$ such that $CH_m C^{-1}=-H_m$ with $m=1,2,3$.  
In momentum space, the  winding number $W_k$ is computed from the anti-diagonal part of the flattened Hamiltonian \cite{Maffei_2018, He_2021}. On the other hand, one has to use the position operator along with flattened Hamiltonian to compute the winding number $W_r$ under OBC, see SM \cite{supp} for more details.  Note that   fermionic operator $l_{j + N} = \pm l_j$,  $l=d$ for LR Kitaev, $l=(a,b)$ for SSH2-Kitaev and $l=(a,b,c)$ for SSH3-Kitaev are $+$ and $-$ signs are for PBC and ABC, respectively. The above sign inversion  is guaranteed for $ k' = \frac{2\pi}{N}(n+\frac{1}{2})$  in ABC as compared to $ k = \frac{2\pi n}{N}$  for PBC.
The possible phase boundaries can be estimated from the vanishing gap $\Delta E_{k/k'}$    
derived from the   momentum space version of the above models under PBC/ABC. Note that in the $N\to \infty$ limit,  $\Delta E_{k}$ and $\Delta E_{k'}$ behave identically yielding the same phase boundaries.   
The phase diagrams obtained from $\Delta E_{k}$ and $W_r$ do not match for a finite-size system when there exist LR features, thus, BBC is not satisfied, see SM \cite{supp} for more details. To investigate the BBC, one can compare  $\Delta E_{k'}$ with $W_r$ over a wide range of parameters, focusing on the long- as well as short-range behavior that we discuss below.



\begin{figure}
    \centering
    \includegraphics[width=1\linewidth]{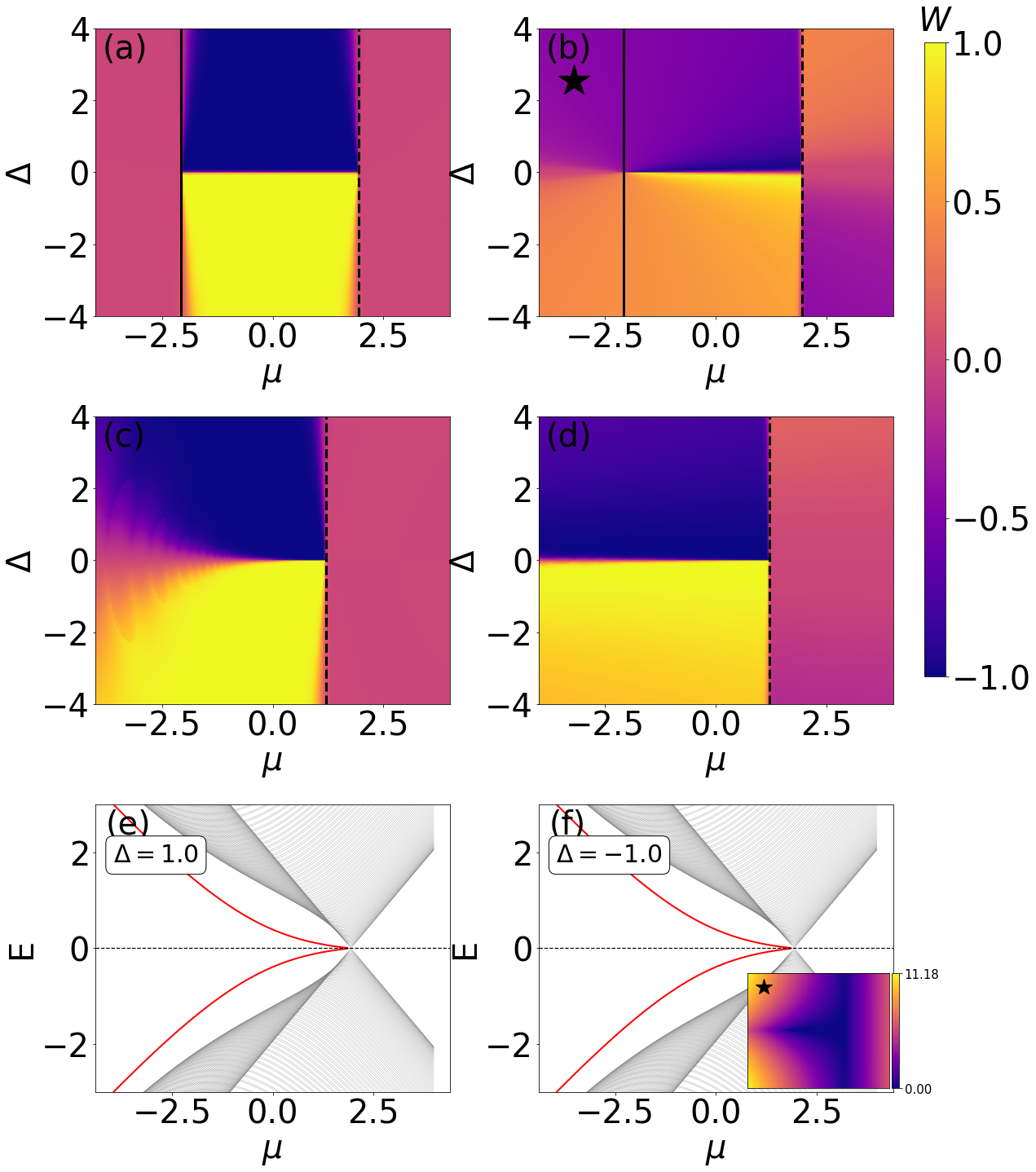}
    \caption{We show the topological phase diagram, obtained from OBC winding number $W_r$, over $\Delta$-$\mu$ plane of power-law Kitaev chain Eq. (\ref{eq:eq1}) for SR hopping and pairing $(\alpha,\epsilon)=(5,5)$ in (a) with $|W_r|=1,0$,  LR pairing $(\alpha,\epsilon)=(0.5,5)$ in (b) with $|W_r|=1/2$, and $<1/2$, LR hopping $(\alpha,\epsilon)=(5,0.5)$ in (c) with $|W_r| \to 1,0$, LR hopping and pairing $(\alpha,\epsilon)=(0.5,0.5)$ in (d) with $|W_r| \to 1,0$.  The black solid [dashed] line represents the analytical solution associated with $k=0$ [$\pi$] mode. (e,f) represent the OBC energy spectrum, exhibiting MDMs, for $\Delta=1$ and $-1$, respectively, with parameters chosen from (b). Here, the system size is $N=100$ and $J=1$.  The inset (f) shows the ABC bulk gap $\Delta E_{k'}$ closing associated with (b).  }
    \label{fig:fig2a}
\end{figure}


\textit{\textcolor{blue}{Half-integer winding number in LR Kitaev model}}--
For LR Kitaev model, we find $ H_1(k) = h_z(k) \sigma_z + h_y(k) \sigma_y $ where $
h_z(k)= -\mu -2J \sum_{k}  x_{\epsilon}^N$ with $x_{\epsilon}^N=  \sum_{r=1}^{N_e} \frac{\cos(kr)}{r^{\epsilon}}$
and $    h_y(k) = \Delta \sum_{k} y_{\alpha}^N$ with $y_{\alpha}^N=
 \sum_{r=1}^{N_e} \frac{\sin(kr)}{r^{\alpha}}$. Considering the thermodynamic limit $N \to \infty$, the
dispersion is found to be $E^{\pm}_{k,\infty} = \pm \sqrt{(\mu+2Jx_\epsilon^\infty(k))^2+(\Delta y_\alpha^\infty(k))^2}$ with 
$x_\epsilon^\infty(k)={\rm Re}[{\rm Li}_{\epsilon}(e^{ik})]$ and $y_\alpha^\infty(k)={\rm Im}[{\rm Li}_{\alpha}(e^{ik})]$ where Li denotes the polylogarithmic function.
Having known the critical modes $k=0,\pi$ for $\Delta\ne 0$ in the Kitaev model, one can find $x_\epsilon^\infty(k\to 0) \simeq \Gamma(1-\epsilon) \sin(\pi \epsilon/2) k^{\epsilon-1} + \zeta(\epsilon) + {\rm O}(k^2) $, 
$x_\epsilon^\infty(k\to \pi) \simeq (2^{1-\epsilon}-1)\zeta(\epsilon) + {\rm O}((\pi-k)^2) $, 
$y_\alpha^\infty(k\to 0) \simeq \Gamma(1-\alpha) \cos(\pi \alpha/2) k^{\alpha-1} + k \zeta(\alpha-1) + {\rm O}(k^3)$, and  $y_\alpha^\infty(k\to \pi) \simeq (\pi-k) (1-2^{1-\alpha}) \zeta(\alpha-1) + {\rm O}((\pi-k)^3)$ where Riemann zeta function $ \zeta(\beta) = \sum_{r=1}^\infty \frac{1}{r^\beta}$ that diverges for $\beta<1$. This leads to the fact that $x_\epsilon^\infty(k\to 0)$ and $y_\alpha^\infty(k\to 0)$ diverge  for $\epsilon<1$ and $\alpha<1$, respectively, otherwise, they converge. On the other hand, $x_\epsilon^\infty(k\to \pi)$ and $y_\alpha^\infty(k\to \pi)$ both converge for $\epsilon>0$ and $\alpha>0$, respectively.  The vanishing band gap $\Delta E^{\infty}_{k}=E^{+}_{k, \infty}-E^{-}_{k, \infty}= \Delta E^{\infty}_{k'}$ determines the phase boundaries of the model that we describe below.

The above analysis ensures that $\mu \to  -2J$ ($2J$) for $\alpha,\epsilon \to \infty$ when $k=0$ ($\pi$), confirming the critical lines, which are depicted by solid (dashed) black lines, of the regular Kitaev model, see Fig. \ref{fig:fig2a}(a). The $\Delta=0$ as well as the above two boundaries are modified as soon as LR superconductivity and/or LR hopping are introduced. For LR   superconductivity [hopping] only, $k=\pi$ boundary is given by $\mu=2J$ [$\mu=2J(1-2^{1-\epsilon})\zeta(\epsilon)$],  see dashed lines in Figs. \ref{fig:fig2a}(b,c).  This phase boundary remains unaltered with $\alpha$ and is
directly obtained from the leading order contribution of $x_\epsilon^\infty(k\to \pi)$.
Interestingly, the $k=0$ boundary vanishes for $\alpha<1$ and/or $\epsilon <1$.   The $\Delta=0$ boundary abruptly ends when $\mu=-2J\zeta(\epsilon) $ for $\epsilon\gg 1$,  see solid lines in Figs. \ref{fig:fig2a}(a,b). The above phase boundary is obtained from the leading term in $x_\epsilon^\infty(k\to 0)$ when $\epsilon>1$. For $\epsilon< 1$, the above phenomenon ceases to exist due to the diverging nature of $\zeta(\epsilon)$.

Having understood the thermodynamic phase boundaries, we now compute $W_r$ from a finite-size system and compare it with the above boundaries in the $\Delta-\mu$ plane for different values of $(\alpha,\epsilon)$. Considering the SR limit with $(\alpha,\epsilon) = (5,5)$ as shown in 
Fig. \ref{fig:fig2a}(a), we find that phase diagram  matches with  that of the conventional $p$-wave Kitaev chain where the phase boundaries appear at $\mu=\pm 2J$ $\forall$ $\Delta$ and $\Delta=0$ for $-2J<\mu<2J$. 
As soon as we introduce only LR superconducting pairing into the system, shown in Fig. \ref{fig:fig2a}(b) for $(\alpha,\epsilon) = (0.5,5)$,
the MZMs, exhibiting $W = \pm 1$ are converted into MDMs, characterized by $W = \pm 0.5$. The critical boundary at $\mu = -2J $ no longer exists, while leaving the $\mu=2J$ boundary unaltered.   By introducing LR hopping only with $(\alpha,\epsilon)=(5,0.5)$ shown in Fig. \ref{fig:fig2a}(c), we find that MZMs, yielding $W = \pm 1$,  persist while
$\mu=2J$ phase boundary shifts to $\mu \simeq J$ and 
the $\mu=- 2J$ boundary is completely dissolved. Interestingly, $\Delta=0$ phase boundary starts disappearing early with $\mu$ in the LR hopping case as compared to the 
LR superconductivity which is consistent with the thermodynamic explanation of phase boundaries.  In Fig. \ref{fig:fig2a}(d) with LR hopping and superconductivity together $(\alpha,\epsilon) = (0.5,0.5)$,  we observe that the $|W_r|$  increases from  half-integer but remains less than unity and there exist one clear phase boundary at $\mu=J$, however, around $\Delta=0$ the phase changes.  Therefore, the LR superconductivity promotes MDMs  and induces topology for $\mu<0$ by resolving $\mu=-2J$ phase boundary. On the other hand, LR hopping causes MZMs to persist and shrinks the topological phases for $\mu >0$ by shifting the phase boundary $\mu < 2 J$ towards $\mu=0$. Importantly, the $\Delta=0$ phase separation is stabilized by a simultaneous effect of LR superconductivity and hopping while retaining the MDMs in the topological phase, see SM \cite{supp} for more details.

The winding number $W_r=1/2$ is accompanied by MDMs that we depict in Figs. \ref{fig:fig2a}(e, f), respectively for $\Delta=1$, and $-1$ with $(\alpha,\epsilon) = (0.5,5)$. One can clearly notice that as soon as the OBC bulk gap closes at $\mu=2J$ and the MDMs, plotted in red color, start appearing which refers to a topological phase transition at $\mu= 2 J$.  
To connect the emergence of MDMs with momentum space gap closing, we demonstrate the ABC bulk gap $\Delta E_{k'}$ for a finite size system in the inset of Fig. \ref{fig:fig2a}(f) where the gapless nature is noticed for $\mu=2 J$ and $\Delta=0$ with $\mu< 2J$. Therefore, the BBC is ensured between boundary modes in real space and ABC bulk gap in the momentum space while   the PBC gap, obtained from  momentum space, fails to identify the critical point accurately in the LR Kitaev model of  finite size.  Intriguingly,  $W_r\simeq \pm 0.5$ for $\mu>2J(1-2^{1-\epsilon})\zeta(\epsilon)$  in Fig. \ref{fig:fig2a}(b) where there exist no mid-gap modes.

\label{s4:model2}
\begin{figure}
    \centering
    \includegraphics[width=1\linewidth]{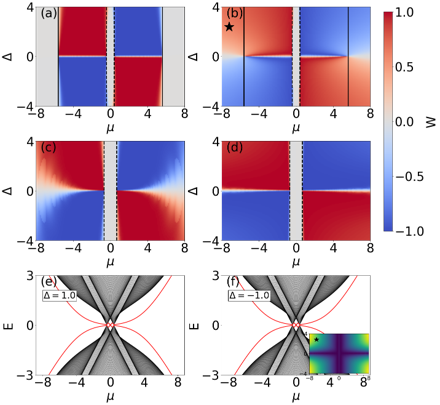}
    \caption{We repeat Fig. \ref{fig:fig2a} for power-law SSH2-Kitaev chain Eq. (\ref{eq:realssh}) with identical set of exponents  
    $(\alpha,\epsilon)$.  In (a,c,d), $|W_r|=1,0$, while in (b), $|W_r|\to 1,0$. Similar to Fig. \ref{fig:fig2a},    the black solid [dashed] line represents the analytical solution at $k=0$ [$\pi$]. 
    The MDMs, corresponding to (b), are depicted in (e,f) for $\Delta=1$, and $-1$, respectively. The inset of (f) shows ABC bulk gap closing of (b) establishing the ABC-OBC correspondence. We consider $N=100$, $V=2.5$, $W_1$=1 and $W_2=2$.}
    \label{fig:fig3a}
\end{figure}

\textit{\textcolor{blue}{Integer winding number in LR SSH2-Kitaev model}}-- Following the Fourier transformation of Eq. (\ref{eq:realssh}), one can obtain the momentum space Hamiltonian in the basis $(a_k,  b_k,  a_{-k}^\dagger,  b_{-k}^\dagger)$ as $H_2(k)= -\mu \sigma_z  \tau_o - g_{\alpha}(k) \sigma_y  \tau_o  +(V+f_{R,\epsilon}) \sigma_z  \tau_x + f_{I,\epsilon} \sigma_z  \tau_y $ with the Pauli matrices $\tau(\sigma)$ operating on the sublattice (spin) degrees of freedom and $f_{R,\epsilon}(f_{I,\epsilon})$ represent the real (imaginary) part of the function $f_{\epsilon}(k)$
where $f_{\epsilon}(k) = (W_1 + W_2) x^N_{\epsilon}  + i(W_2 - W_1) y^N_{\epsilon} $ and $g_{\alpha}(k) = 2i\Delta y^N_{\alpha} $.  In the thermodynamic limit $N\to \infty$, the band dispersion is found to be $E^{\pm}_{k,\infty} = \pm \big( (g^{\infty}_{\alpha})^2 +(-\mu\pm ((V+f^{\infty}_{R,\epsilon})^2+(f^{\infty}_{I,\epsilon})^2)^{1/2} )^2\big)^{1/2}$. The gapless conditions are as follows:  $g^{\infty}_{\alpha}=0$ and $\mu =\pm \sqrt{(V+f^{\infty}_{R,\epsilon})^2+(f^{\infty}_{I,\epsilon})^2}$. 
In the SR limit, $k=0$ and $\pi$ boundaries are given by  $\mu \simeq \pm(V+W_1+W_2) $, depicted by solid lines, and $\mu \simeq \pm(V-W_1-W_2) $,  depicted by dashed lines, respectively, and they match exactly with those of the nearest neighbour model, see Fig. \ref{fig:fig3a}(a).  
In the LR limit, as discussed in the previous model, $k=0$ critical boundaries vanish for $\alpha,\epsilon<1$ due to the divergences in $x,y$ terms similar to the previous model. As a result, $\mu = \pm(V+W_1+W_2)$ boundary vanishes. On the other hand, $k=\pi$ critical boundaries survive in the above limit leading to $    \mu = \pm [V-(W_1+W_2)(1-2^{1-\epsilon}\zeta(\epsilon)]$ as depicted by black dashed line in Figs. \ref{fig:fig3a}(b,c,d). The $\Delta=0$ boundary abruptly ends at $\mu =\pm [V + (W_1 +W_2)\zeta(\epsilon)$, depicted by black solid line in Fig. \ref{fig:fig3a}(b), that is caused by leading contribution in $x_\epsilon^\infty(k\to 0)$ when $\epsilon>1$. Similar to the earlier model, for $\epsilon<1$, such sudden termination of $\Delta=0$ line does not take place owing to the divergence in $\zeta(\epsilon)$. Therefore, the open-ended critical line at $\Delta=0$ is a generic feature of such LR models with a divergence associated with $k\simeq 0$ modes irrespective of the number of sub-lattice.

Now coming to the winding number profile of a finite-size SSH2-Kitaev model , we demonstrate the phase diagram  in $\Delta-\mu$ plane where we observe $|W_r|=1$ or $0.5 \ll |W_r| \simeq 1$ in the topological phase irrespective of the choice of $\alpha$ and $\epsilon$. In the SR limit, there exist MZMs for $|V-W_1-W_2|<|\mu|< |V+W_1+W_2|$, see Fig. \ref{fig:fig3a}(a). One can 
interpret the above phase diagram  as the doubling of the topological phases in the conventional Kitaev model over a wider range of $\mu$ while their profile with $\Delta$ alters. 
For LR superconductivity,  MDMs show up while the $|W_r|$ gradually decreases from unity as $|\mu|$ increases above  $|V-W_1-W_2|$, see Fig. \ref{fig:fig3a}(b). 
$|W_r|$ eventually vanishes when  $|\mu|$ crosses  $|V+W_1+W_2|$ i.e., above the open-ended critical line at $\Delta=0$.   
This type of crossover behavior can also be seen for LR hopping around $\Delta=0$ while the open-ended critical line at $\Delta=0$ is no longer present, see Fig. \ref{fig:fig3a}(c). $|W_r|$ stays close to unity  over a wider parameter space for LR hopping as compared to the LR superconductivity. Interestingly, the MDMs  transform into MZMs when superconducting pairing  is converted from LR to SR, even though the 
LR hopping is present.  However, the trivial region expands here more as compared to the LR superconductivity.  The topological trivial crossover region is significantly suppressed under the simultaneous presence of  LR hopping and superconductivity,  see Fig. \ref{fig:fig3a}(d). Note that  MDMs continue to survive as long as LR superconductivity is present, irrespective of the hopping profile, see SM \cite{supp} for more details.

We show the emergence of MDMs in Figs. \ref{fig:fig3a}(e,f) for $\Delta=\pm 1$, respectively, for LR superconductivity only where the bulk gap-closing is clearly visible for $ \mu = \pm [V-(W_1+W_2)(1-2^{1-\epsilon}\zeta(\epsilon)]$. The inset of Fig. \ref{fig:fig3a} (f) shows the closing of ABC gap $\Delta E_{k'}$, obtained from $H_2(k)$, that  is in exact agreement with the OBC
winding number shown in Fig. \ref{fig:fig3a}  (b). This establishes the  ABC-OBC correspondence for the even number of sublattices yielding nearly integer winding number. One finds mid-gap states inside the topologically trivial region, bounded by $ -[V-(W_1+W_2)(1-2^{1-\epsilon}\zeta(\epsilon)] < \mu   [V-(W_1+W_2)(1-2^{1-\epsilon}\zeta(\epsilon)]$,  with winding number $W_r=0$.

\begin{figure}
    \centering
    \includegraphics[width=1\linewidth]{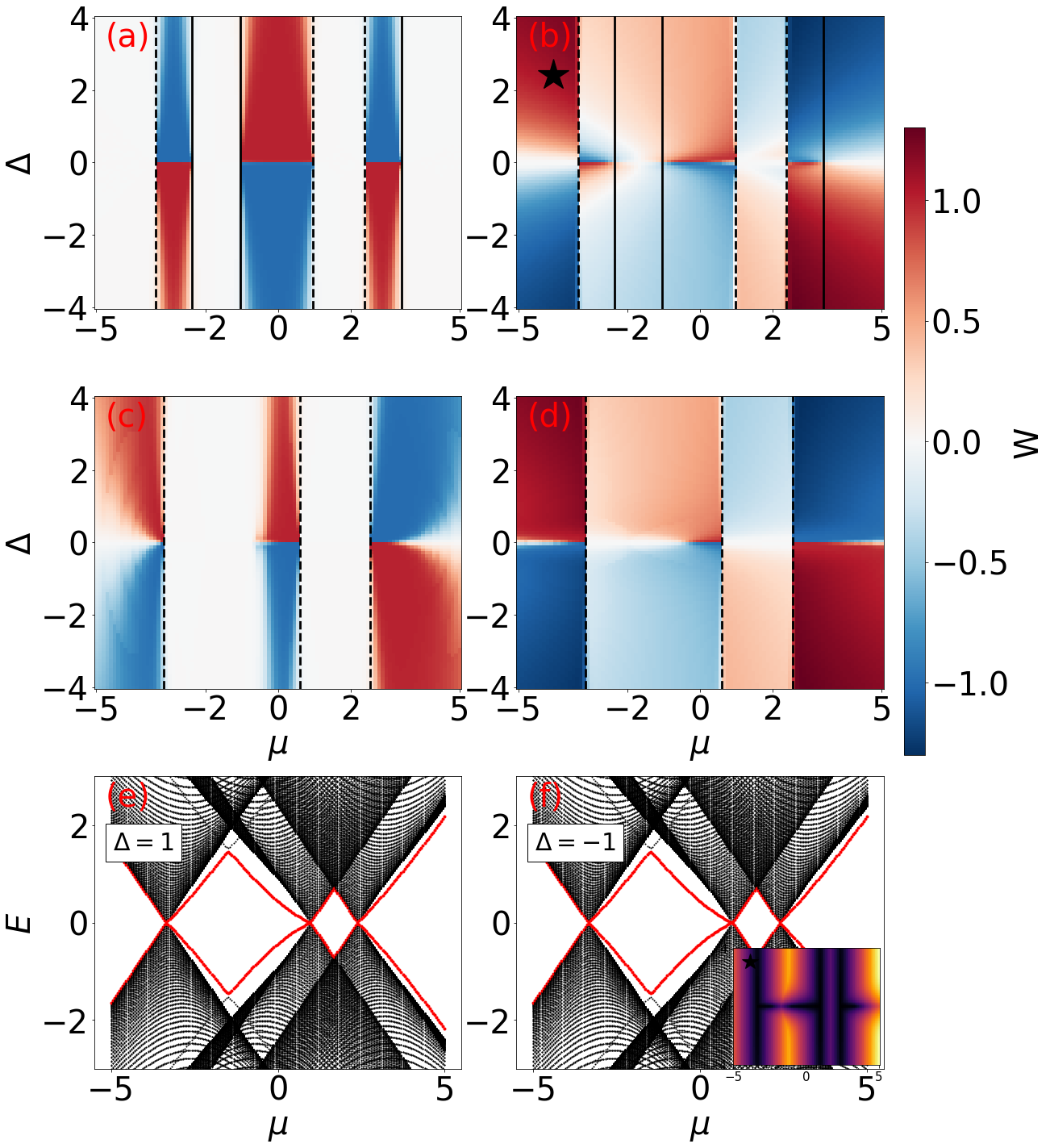}
    \caption{We repeat Fig. \ref{fig:fig2a} for power-law SSH3-Kitaev chain Eq. (\ref{eq:realssh3}) with identical set of exponents  
    $(\alpha,\epsilon)$.  In (a,c), $|W_r|=1,0$, while in (b,d), $|W_r|\to 3/2,1/2,0$. Similar to Fig. \ref{fig:fig2a},    the black solid (dashed) line represents the analytical solution at $k=0$ $(\pi)$. 
    The MDMs, corresponding to (b), are depicted in (e,f) for $\Delta=1$, and $-1$, respectively. The inset of (f) shows ABC bulk gap closing of (b) establishing the ABC-OBC correspondence.  We consider $N=100$, $V_1=2$, $V_2=2$ and $W=1$. }
    \label{fig:fig4a}
\end{figure}

\textit{\textcolor{blue}{Half-integer winding number in LR SSH3-Kitaev model}}-- One can construct the momentum space version of LR SSH3-Kitaev model using 
Eq. (\ref{eq:realssh3}), as given by 
\begin{equation}
H_3(k) = \begin{pmatrix} 
H_e(k)-\mu &  \Delta(k) \\
\Delta^{\dagger}(k) & -H^T_e(-k) +\mu  \\ 
\end{pmatrix}
\end{equation}
where the electronic part $H_e(k)$ is written as 
\begin{equation}
H_e(k) = \begin{pmatrix} 
0& V_1 & f_{\epsilon}(k) \\
V_1 & 0 & V_2   \\
f^*_{\epsilon}(k) & V_2 & 0 \\ 
\end{pmatrix}
\end{equation}
and superconductivity part $\Delta(k)$ is given by 
\begin{equation}
\Delta(k) = \begin{pmatrix} 
 g_{\alpha}(k) & 0  &  0\\
 0  &  g(k) & 0 \\
 0  &  0 & g_{\alpha}(k)\\
\end{pmatrix}.
\end{equation}
Here $f_{\epsilon}(k) = W x^N_{\epsilon} - iWy^N_{\epsilon}$ and $g_{\alpha}(k) = 2i\Delta y^N_{\alpha}$. In the thermodynamic limit $N\to \infty$, 
$E^{\pm}_{k,\infty}=\pm \sqrt{|\lambda^{\infty}_{\epsilon}(k)|^2+|g^{\infty}_{\alpha}(k)|^2}$ where  $\lambda^{\infty}_{\epsilon}(k)$ is obtained from the $\det[H_e(k)-\mu]=0$. The critical phenomena is determined by  $\lambda^{\infty}_{\epsilon}(k)=-\mu^3 +\mu(V_1^2+V_2^2+|f^{\infty}_{\epsilon}(k)|^2) + V_1V_2[f^{\infty}_{\epsilon}(k)+(f^{\infty}_{\epsilon}(k))^*] =0$ and $g^{\infty}_{\alpha}(k)=0$.
In the SR limit $\alpha, \epsilon\gg1$,  $k=0,\pi$ both modes determine phase boundaries. The critical boundary is obtained from the roots of the cubic equation $\mu^3-a \mu-b = 0$ yielding $\mu_n = 2\sqrt{a/3} \cos[(1/3)\cos^{-1}(3\sqrt{3}b/2a^{3/2})-2\pi n/3]$ with $n=1,2,3$ that accurately produce the nearest neighbour results. To be precise,  the $k=0$ boundary, designated by black solid line, is determined by  $a=V_1^2+V_2^2+W^2$ and $b=2 V_1 V_2 W$ while $k=\pi$ boundary, depicted by a dashed black line,  is determined by  $a=V_1^2+V_2^2+W^2$ and $b=-2 V_1 V_2 W$. These boundaries are in excellent agreement with the numerical phase diagram of $W_r$ as shown in Fig. \ref{fig:fig4a} (a). 
As discussed earlier, for LR limit $\alpha, \epsilon<1$, the critical behavior is caused by 
$\lambda^{\infty}_{\epsilon}(k\to \pi)= 0$ with $a=V_{1}^{2} +V_{2}^{2}+W^{2}\left(1-2^{1-\epsilon}\right)^{2}\zeta^{2}(\epsilon)$ and $b=-2V_{1}V_{2}W\left(1-2^{1-\epsilon}\right)\zeta(\epsilon)$.
The resulting boundaries correctly produce the boundary obtained numerically as depicted by black dashed line in Figs.  \ref{fig:fig4a} (b,c,d).  The open-ended $\Delta=0$ critical line, as found for $\epsilon>1$,
abruptly ends when $\lambda^{\infty}_{\epsilon}(k\to 0)$ with $a=V_1^2+V_2^2+W^2\zeta^2(\epsilon)$ and $b= 2V_1V_2W\zeta(\epsilon)$, see Fig.  \ref{fig:fig4a} (b). The open-ended $\Delta=0$ critical line no longer exists  for  $\epsilon>1$ similar to the earlier models.

We now discuss the winding number phase diagram for a finite-size SSH3-Kitaev model in Fig. \ref{fig:fig4a} where $|W_r| \simeq 1/2$, $1$ $3/2$ within the topological phase in the presence of LR superconductivity. In the SR limit with $\alpha, \epsilon =5$, we find topological phases with $|W_r|=1$, hosting MZMs, similar to that of the nearest neighbour model, Fig. \ref{fig:fig4a} (a).  One can  interpret the above phase diagram  as the tripling of the topological phases in the conventional Kitaev model over a wider range of $\mu$. 
Upon introducing a LR superconducting pairing term $ (\alpha=0.5,\epsilon=5)$, as shown in Fig. \ref{fig:fig4a}(b), we observe $|W_r| \simeq 3/2$ for $\mu>2.38$ and $\mu<-3.35$ while MDMs are observed only for $\mu>2.38$.  There exist MDMs between $-3.35<\mu<0.97$, while $|W_r|$ varies from $\simeq 1/2$ to $<1/2$. The open-ended $\Delta=0$ critical line is present within the above window, bearing resemblance to the LR Kitaev model.   In Fig. \ref{fig:fig4a}(c) with LR hopping only, we find $|W_r|=1$ in the topological phase, hosting MZMs only. This behavior is qualitatively similar to that of a LR Kitaev model discussed previously. In the simultaneous presence of LR hopping and superconductivity shown in Fig. \ref{fig:fig4a}(d), we find $|W_r| \simeq 1/2$ and $3/2$ along with the MDMs.

To highlight the emergence of MDMs in this model, we show OBC spectrum associated with $\alpha=0.5$, $\epsilon=5$ for $\Delta=1$ and $-1$ in Figs. \ref{fig:fig4a}(e,f), respectively. The change in $W_r$ profile in the LR limit is accurately captured by  closing of ABC gap $\Delta E_{k'}$, derived from momentum space, indicating the ABC-OBC correspondence, see the inset of Fig. \ref{fig:fig4a}(f), see SM \cite{supp} for more details.

\textit{\textcolor{blue}{Conclusion}}--
In this work, we studied the topological properties of LR Kitaev chain and its higher sub-lattice analogs by suitably incorporating SSH-type model. Considering power-law profile of hopping and superconductivity, we find that zero-momentum  (finite-momentum) critical phenomena perish (survive) under the LR hopping and/or LR superconductivity, while the zero-momentum mode causes an open-ended critical line to emerge with a vanishing superconducting gap for LR superconductivity. Importantly, the critical boundaries are governed solely by the hopping exponent. The open-ended critical line vanishes if  LR hopping is only present, leaving the finite-momentum critical phenomena. 
As a result, the zero-momentum critical phenomena are only present (completely absent) in the universal SR (LR) limit where hopping and superconductivity are both SR (LR).

Having known the above common features for all the models in the thermodynamic limit, we uncover a strange BBC in the finite-size system. The emergence of boundary modes under OBC in real space is connected with the gap closing under ABC  in momentum  space. 
Such a feature in the  LR limit nicely translates into OBC-PBC correspondence in the SR limit. The model supports MDMs in the LR limit while MZMs in the SR limit. Most intriguingly, in the case of LR superconductivity only, we find half-integer winding numbers for models with an odd number of sub-lattices while integer winding number for even number of sub-lattices. These half-integer winding numbers reduce to integers in the SR limit of superconductivity. Surprisingly, LR hopping promotes MZMs rather than MDMs, rendering an interesting profile of mid-gap states due to the interplay between power-law hopping and superconductivity. However, there is apparently no correlation between half-integer winding number and MDMs as the later continues to exist in even sub-lattice model where winding number is found to be integer only.

\textcolor{blue}{\textit{Acknowledgement}}--
We sincerely thank Archak Purkayastha for  useful discussions on the energy of the power-law models.   TN acknowledges the
NFSG ``NFSG/HYD/2023/H0911" from BITS Pilani.  TN thanks the
Advanced Research Grant (ARG) from Anusandhan National Research Foundation Grant No.
ANRF/ARG/2025/003163/PS.

\bibliography{bibfile}{}

\end{document}